\documentclass[journal]{IEEEtran}

\ifCLASSINFOpdf
\else
   \usepackage[dvips]{graphicx}
\fi
\usepackage{url}

\usepackage{graphicx}
\usepackage{multirow,array}
\usepackage{hhline}
\usepackage{amsmath}
\usepackage{amssymb}
\usepackage{rotating}
\usepackage{xcolor}
\usepackage{balance}

\begin{document}
\title{Adaptive Reverberation Absorption using Non-stationary Masking Components Detection for Intelligibility {\color{black} Improvement} }

\author{G. Zucatelli, \IEEEmembership{Student Member, IEEE}, and R. Coelho, \IEEEmembership{Senior Member, IEEE}
\thanks{The authors are with the Laboratory of Acoustic Signal Processing (lasp.ime.eb.br), Military Institute of Engineering (IME), Rio de Janeiro, Brazil (e-mail: coelho@ime.eb.br). This work was partilly supported by the National Council for Scientific and Technological Development (CNPq) 307866/2015 and Funda\c{c}\~ao de Amparo à Pesquisa do Estado do Rio de Janeiro (FAPERJ) 203075/2016. This work is also supported by the Coordena\c{c}\~ao de Aperfei\c{c}oamento de Pessoal de N\'ivel Superior - Brasil (CAPES) - Grant Code 001.}

}

\maketitle

\begin{abstract}
This letter proposes a new time domain absorption approach designed to reduce masking components of speech signals under noisy-reverberant conditions.
In this method, the non-stationarity of corrupted signal segments is used to detect masking distortions based on a defined threshold.
The non-stationarity is objectively measured and is also adopted to determine the absorption procedure.
Additionally, no prior knowledge of speech statistics or room information is required for this technique.
{\color{black} Two intelligibility measures (ESII and ASII\textsubscript{ST}) are used for objective evaluation.
The results show that the proposed scheme leads to a higher intelligibility improvement when compared to competing methods.
A perceptual listening test is further considered and corroborates these results.  
Furthermore, the updated version of the SRMR quality measure (SRMR\textsubscript{norm}) demonstrates that the proposed technique also attains quality improvement.
} 


\end{abstract}


\IEEEpeerreviewmaketitle

\vspace{-.6 cm}
\section{Introduction}

\IEEEPARstart{S}{peech} communication commonly takes place in enclosed and urban environments such as concert halls, kitchens and offices.
Along with the direct acoustic signal propagation between source and listener locations, the sound reverberates due to reflection in walls and surfaces.
While the early reflections (ER) can improve speech intelligibility, late reverberation (LR) may cause quality and intelligibility reduction \cite{Bolt_1949}\cite{Nabelek_1993}\cite{Assmann_2004}\cite{Brad_03}.

Room impulse response (RIR) typically describes the sound propagation and is generally described by the reverberation time ($T_{60}$) and the direct-to-reverberant ratio (DRR).
Speech signals can also be degraded by background acoustic noises (Babble, Chainsaw and Cafeteria) present in the urban space.
Such effects are non-stationary masking components and represent a major drawback to speech intelligibility improvement.

In the literature, speech enhancement solutions were designed to cope with background non-stationary noises \cite{Gerk_12}\cite{Zao_14}\cite{Zao_15}\cite{Tavares_16} attaining interesting results for quality and intelligibility. 
However, room reverberation is not considered by these techniques.  
Adaptive time-domain pre-processing methods were proposed to improve speech intelligibility by mitigating the reverberation effect.
The Steady State Suppresion (SSS) \cite{Hodo_10} solution considers the importance of transient regions of speech for intelligibility and suppresses steady-state frames to reduce overlap masking effects. 
A more recent approach, the Adaptive Gain Control (AGC) \cite{Petkov_16} method uses prior knowledge of speech statistics and the RIR information to adaptively improve or reduce the energy of speech frames.
Both methods operate prior to speech signal presentation in a room, such that the resulting reverberated signal is similar to its anechoic version.

This letter proposes a new time-domain method denominated Adaptive Reverberation Absorption with Non-Stationary Detection (ARA\textsubscript{NSD}). 
Different from SSS and AGC techniques, {\color{black} the main idea of this proposal is to act similar to a physical element, changing the low absorption characteristic of materials that compose a room in the listener position}.
{\color{black} One major advantage of ARA\textsubscript{NSD} is that it adaptively absorbs masking components of corrupted speech signals.
Thus, leading to speech intelligibility improvement with no prior knowledge of the RIR or speech statistics}.
The Index of Non-Stationarity (INS) \cite{Flandrin_10} is selected as an objective measure for the detection of masking components.
A non-stationarity threshold is defined for the proposed frame-by-frame absorption procedure.

Extensive experiments are conducted to objectively evaluate the ARA\textsubscript{NSD} method for speech intelligibility {\color{black} improvement}.
The noisy-reverberant scenario is composed of two real reverberant rooms and four background non-stationary acoustic noises with five different SNR values.
The ESII \cite{Rheber_05} and ASII\textsubscript{ST} \cite{Hen_15} measures are adopted for the intelligibility prediction. 
These measures are explicitly designed to deal with the non-stationarity of speech and its distortions.
The SRMR\textsubscript{norm} \cite{Falk_14} measure is further considered as it is primarily used for signals under reverberation effect.
{\color{black} A subjective listening test is also performed and results show that the proposed method outperforms the competing techniques in terms of speech intelligibility}.

\vspace{-.4cm}  
\section{Reverberation and Non-Stationarity}
\vspace{-.1cm} 

The reverberation effect is usually defined as a linear filtering process such that, given a RIR $h(n)$, the reverberated signal can be obtained by convolution. 
In real environments, acoustic noises are also a common distortion, which means that the resultant noisy-reverberant speech signal $s(n)$ can be obtained by $s(n) = x(n) \ast h(n) + w(n)$, where $x(n)$ is the clean speech signal and $w(n)$ is the background noise.

The Index of Non-Stationarity (INS) \cite{Flandrin_10} is here defined to objectively examine the non-stationarity of speech signals under noisy-reverberant environments.
This measure compares the target signal with stationarity references called surrogates for different time scales $T_h/T$, where $T_h$ is the short-time spectral analysis length and $T$ is the total signal duration.
For each length $T_h$, a threshold $\gamma$ is defined to keep the stationarity assumption considering a $95\%$ confidence degree as
\begin{equation}
 INS \quad \left\{\begin{matrix}
 \hspace{-.78cm} \leq \gamma, \quad \text{signal is stationary} \\
 >  \gamma, \quad \text{signal is non-stationary}. 
\end{matrix}\right.
\end{equation}

Figure \ref{Fig::INS_p1} illustrates the spectrograms and INS values obtained for a direct speech signal and its corresponding reverberated version in the Aula Carolina\footnote{RIR collected from the AIR database \cite{Air_09}.} room with $T_{60}=4.9$ s in two conditions: without and with a background Chainsaw noise at $-3$ dB.
Note that reverberation and acoustic noise significantly change the temporal and spectral structure of speech signal. {\color{black} These masking effects can engender intelligibility reduction \cite{Bolt_1949}\cite{Nabelek_1993}\cite{Assmann_2004}.
Furthermore, the non-stationary behavior of the natural speech signal is considerably attenuated, varying its maximum INS value from 200 to around 100}.
The background Chainsaw noise increases the INS value in small scales, which means that short-time segments become more distinct from the overall signal. 
{\color{black} As INS alters on noisy-reverberant scenarios, it can be a useful instrument for detection of such effects. 
In this work, the INS is adopted for detection and reduction of masking components.}

\vspace{-.3cm}
\section{Adaptive Reverberation Absorption with Non-Stationary Detection}
\label{Sec::Method}
\vspace{-.1cm}
The ARA\textsubscript{SND} method is presented in this section. 
The technique is described in two main phases: reverberation detection and acoustic absorption.  

\vspace{-.3cm}
\subsection{Reverberation Detection}
\vspace{-.1cm}
\label{Sub::Detect}

A reverberation group (RG), denoted as $s_{RG} (m,n)$, is here defined as the $m$-th segment composed of $N=8$ consecutive frames of the corrupted speech.
This window duration is selected to enable a long-term temporal observation of the reverberation effect and detect noisy-reverberant masking components.
Successive RGs are obtained considering a $50$\% overlap between signals.

\begin{figure}[t!]
  \begin{center}
  \centering
  \includegraphics[width=8.0cm]{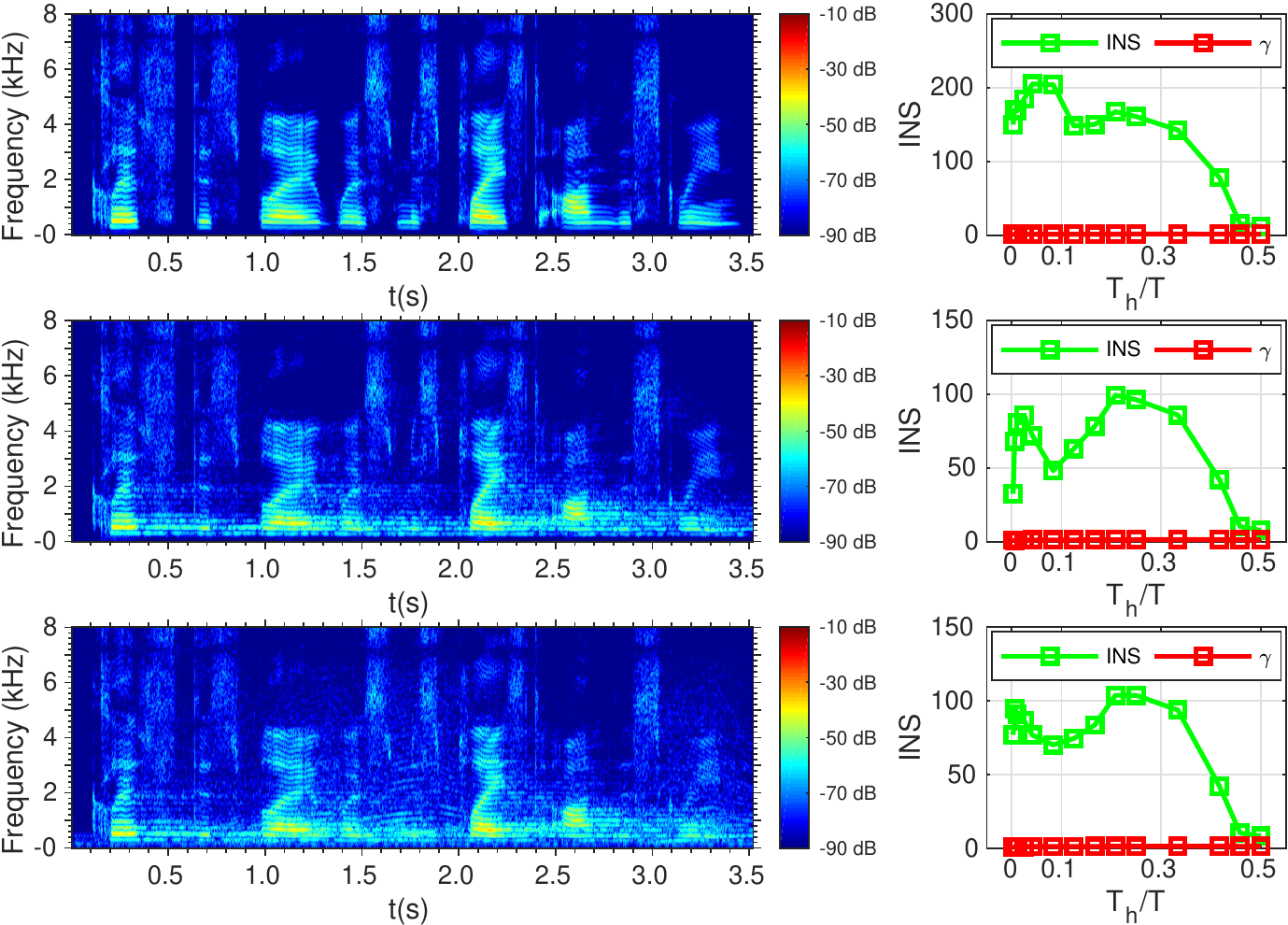}
  \vspace{-.3cm}
  \caption{Spectrogram and related INS for direct signal (top), reverberated signal with $T_{60} = 4.9s$ and SRR $= 7.1$ dB (middle), and reverberated signal with Chainsaw noise at $-3$ dB (bottom).  }
  \vspace{-.2cm}
  \label{Fig::INS_p1}    		
  \end{center}
\end{figure}

\begin{figure}[t!]
  \begin{center}
  \centering
  \vspace{-.3cm}
  \includegraphics[width=8.5cm]{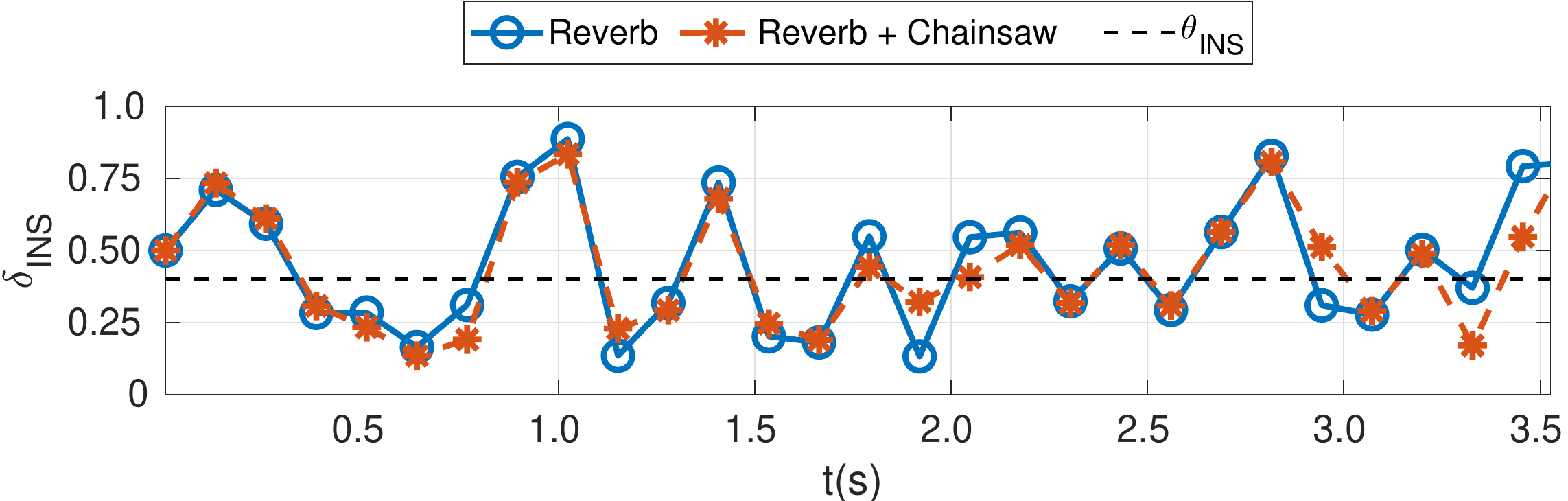}
  \vspace{-.4cm}
  \caption{Non-stationarity variation ($\delta$\textsubscript{INS}) for the reverberated and noisy-reverberated signals.}
  \vspace{-.95cm}
  \label{Fig::Dins}    		
  \end{center}
\end{figure}

For each $s_{RG} (m,n)$, the INS values are computed considering different scales of $T_h/T$.
The INS values obtained for all scales are grouped into a vector ${\bf v_{INS}}(m)$ which characterizes the non-stationary behavior of the $m$-th RG. 
Consecutive vectors are then used to compute a normalized variation of the non-stationary property as
\begin{equation}
 \delta_{INS}(m) = \frac{||{\bf v_{INS}}(m)-{\bf v_{INS}}(m-1)||}{||{\bf v_{INS}}(m)||+||{\bf v_{INS}}(m-1)||} .
 \label{Eq::Delta}
\end{equation}
Figure \ref{Fig::Dins} shows the $\delta_{INS}$ values obtained for the reverberated and noisy-reverberated speech signals of Figure \ref{Fig::INS_p1}. 
Note that even with masking components, important speech regions, e.g. the ones near  $0.2$ s, $1.1$ s, $1.4$ s and $2.7$ s (refer to Fig.1 (top)), are still identified by the highest values of $\delta_{INS}$ in both conditions.	 
Moreover, masked regions closed to $0.7$ s, $1.9$ s and $3.0$ s attain low $\delta_{INS}$ values.
This demonstrates that the proposed $\delta_{INS}$ is an interesting detection approach for noise and reverberation masking components.
The $\theta_{INS}$ (black dashed line) in Figure \ref{Fig::Dins} illustrates a threshold of non-stationarity defined by the median value of $\delta_{INS}$.
In this example, $\theta_{INS}$ value is $0.4$ indicating the difference of the speech and noisy-reverberant regions.


\vspace{-.2cm}
\subsection{Acoustic Absorption}
\vspace{-.1cm}
The proposed ARA\textsubscript{NSD} absorption approach is implemented on a frame-by-frame basis and is established depending on the value of $\theta_{INS}$.
For each frame $s_{frm}(l,n)$, a INS vector ${\bf v_{frm}}(l)$ is extracted similarly as in Section \ref{Sub::Detect}.
A short-time distance $d(l) \in [0,1]$ is then computed as in (\ref{Eq::Delta}) and used to determine the $l$-th frame absorption.


Sigmoid functions are selected to assign each value of $d(l)$ to a corresponding absorption $A(m,l)$ because of their smoothness and monotonic property.
The proposed adaptive absorption $A(m,l)$ is therefore defined in every frame $l$ by
\vspace{-.1cm}
\begin{equation} 
\resizebox{.44 \textwidth}{!}
{
    \hspace{-.35cm}$A(m,l) = \left\{\begin{matrix}
      F(l). \frac{L(m) - S}{1 + \exp(-k.(d(l)-d_0))} + S \text{, } \delta_{INS} \leq \theta_{INS}; \\
    \frac{L'}{1 + \exp( -k'.(d(l) - d'_0 )} \hspace{1.5cm} \text{, } \delta_{INS} > \theta_{INS}, 
    
    \end{matrix}\right.$
}
\end{equation}
where $d_0$ and $d'_0$ are the inflection points with corresponding growth rate of $k$ and $k'$.
The $S$ stands for a minimum shift in order to avoid total absorption of signal frames. 
Moreover $L(m)$ and $L'$ are the maximum absorption values.
As the noisy-reverberant masking effect is non-stationary by nature it is important to determine an adaptive upper bound absorption.
Both $L(m)$ and the $F(l)$ factor are considered for this task.
The first one is updated accounting the overlapped region as $L(m) = p\delta_{INS} + (1-p)L(m-1)$, where $p$ assigns the importance of the present RG signal. 
The second term is defined as the factor $F(l) = d(l)^{1.2-d(l)}$ to guarantee that $A(m,l) \approx L(m)$ only for $d(l) \approx 1$.  
As $d(l)$ represents the short-term non-stationarity behavior, the absorption maintain a high value if $d(l) \approx 1$ for it refers to an important speech region.
The processed signal $s'(n)$ is obtained by overlap add process of absorbed frames $s'_{frm}(l,n) = A(m,l).s_{frm}(l,n)$.  

\begin{figure}[t!]
  \vspace{-.3cm}
  \begin{center}
  \centering
  \includegraphics[width=8.0cm]{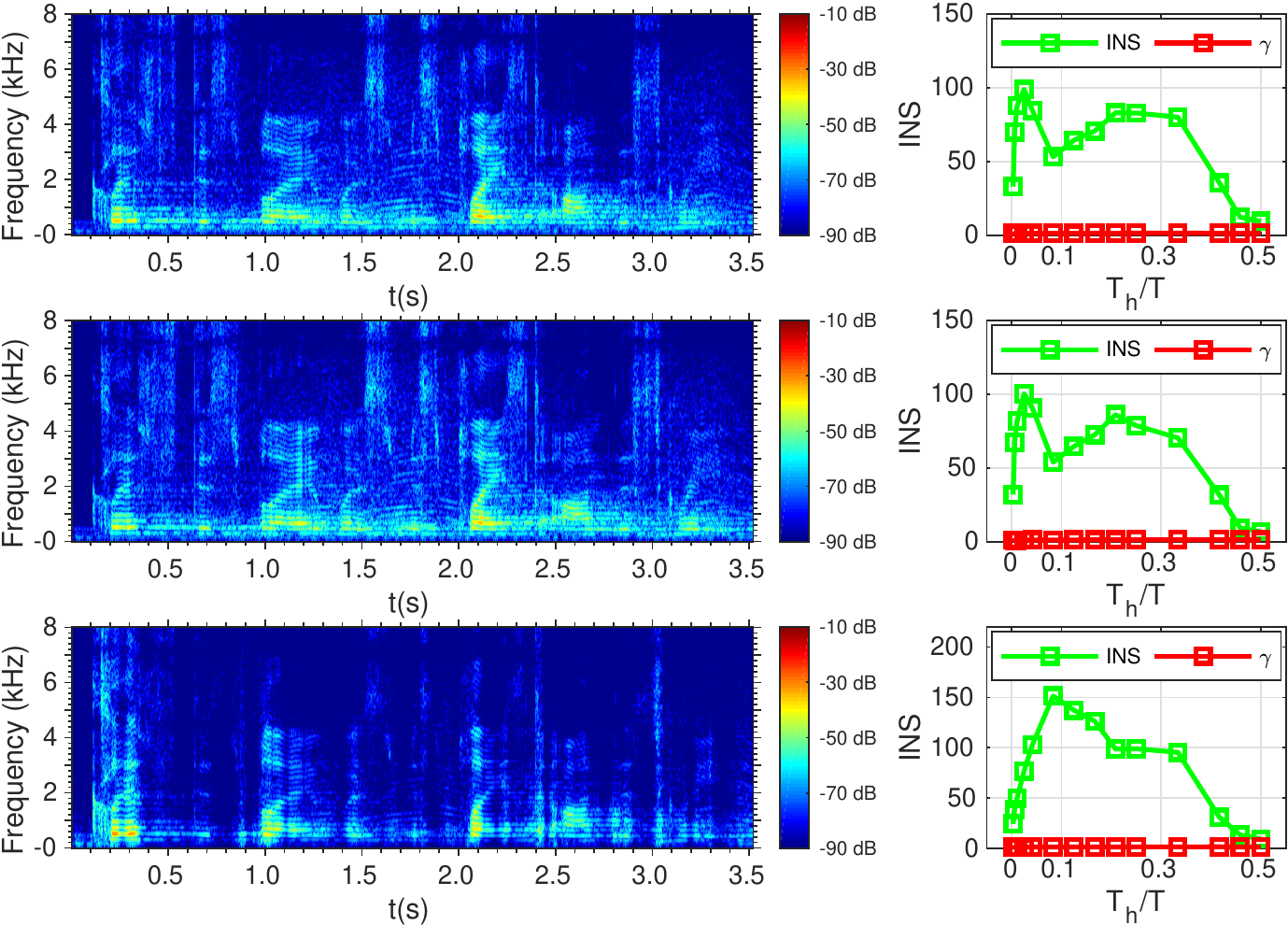}
  \vspace{-.3cm}
  \caption{Spectrogram and related INS for noisy-reverberant signal processed by  SSS (top), AGC (middle) and ARA\textsubscript{NSD} (bottom).  }
  \vspace{-.7cm}
  \label{Fig::INS_p2}    		
  \end{center}
\end{figure}

Figure \ref{Fig::INS_p2} depicts the spectrograms and INS values of the noisy-reverberant signal (refer to Figure \ref{Fig::INS_p1} (bottom)) processed by the baseline SSS, the AGC technique and the proposed ARA\textsubscript{NSD}.
Note that, the ARA\textsubscript{NSD} is able to absorb masking components of the corrupted signal, e.g. near $0.7$s and $1.5$s, which makes the resulting signal more similar to its anechoic version.
Moreover, the proposed method restores the natural non-stationarity behavior raising the INS value from $100$ up to $150$, which is closer to the direct signal. 

\vspace{-.4cm}  
\section{Experiments and Discussion}
\vspace{-.1cm} 
Several noisy-reverberant conditions are used to evaluate the SSS \cite{Hodo_10}, the AGC \cite{Petkov_16} and the proposed ARA\textsubscript{NSD} technique in terms of intelligibility.
A subset of 24 speakers (16 male and 8 female) are randomly selected from the TIMIT speech database \cite{Timit_93}, which leads to a total of 240 speech signals (ten for each speaker). 
>From these, 100 are arbitrarily chosen for the test signal and the remaining are used on the speech modeling step of AGC.
Each speech segment is sampled at $16$ kHz and has, on average, 3 seconds.
Two real reverberation rooms from the AIR database \cite{Air_09} are considered in the experiments.
The Stairway is characterized by a medium reverberation time ($T_{60} = 1.1$ s) and a small value of DRR $= -9.1$ dB.
The Aula Carolina room presents parameters of $T_{60} = 4.9$ s and a higher value of DRR $= 7.1$ dB.
Both RIRs are equalized for a total energy of $17.9$ dB.
The Babble, SSN, Cafeteria and Chainsaw additive background noises are selected, respectively, from the RSG-10 \cite{Steen_88}, DEMAND \cite{Thie_2013} and Freesound.org\footnote{Available at www.freesound.org.} databases.
Except for the SSN, all other noises are characterized with non-stationary behavior.
  
Speech signals are corrupted considering five SNRs values varying from $-3$ dB up to $1$ dB, where the SNRs are measured between the original unprocessed speech and the background noise.
{\color{black} The SNR range is adopted to guarantee ESII and ASII\textsubscript{ST} scores between $0.45$ and $0.75$ for the unprocessed (UNP) speech signal in all scenarios.
These values are defined as thresholds of poor and good intelligibility \cite{ANSI_97}\cite{Sauert_06}, respectively. 
All UNP intelligibility scores are presented in Tables \ref{Tab::ESII_UNP} and \ref{Tab::ASII_UNP}.
The smallest value (ESII$=0.48$) is achieved for the Stairway room with the highly non-stationary Chainsaw noise at $-3$ dB. 
The Aula Carolina room with Cafeteria noise at $1$ dB presents the highest score of ASII\textsubscript{ST}$=0.71$.}
The ARA\textsubscript{SND} operates with $32$ ms frames and $\theta_{INS}=0.4$.
The maximum value for relevant speech regions $L'$ is set to $1.2$ and the RG importance to $p=0.7$ in all scenarios. 
The sigmoid parameters are fixed to $k=17$ for $d=-0.2$ and $k'=13$ for $d'=0.5$.
The minimum shift $S$ is set to $0.05$.

\vspace{-.4cm}
\subsection{Objective Evaluation of Intelligibility}
\vspace{-.1cm}
\begin{table}[t!]
\vspace{-.3cm}
  \caption{ESII intelligibility measure [\%] for UNP speech signals}
  \newcommand{\mc}[3]{\multicolumn{#1}{#2}{#3}}
  \centering
  \renewcommand{\tabcolsep}{1.0mm}
  \vspace{-.2cm}
\begin{tabular}{|c|l c c c c c c c c c c c|} 	
   \hline
   \multicolumn{2}{|l}{ \bf  }    & \multicolumn{5}{c}{ Stairway ($T_{60} = 1.1$ s) } & & \multicolumn{5}{c|}{ Aula Carolina ($T_{60} = 4.9$ s) } \\ \hhline{~~-----~-----} 
   \multicolumn{2}{|l}{ SNR (dB) }    & {\centering -3} & {\centering -2} & {\centering -1} & {\centering 0} & {\centering 1} & & {\centering -3} & {\centering -2} & {\centering -1} & {\centering 0} & {\centering 1} \\ \hline \hline
   \centering  \multirow{4}{*}{\begin{sideways} {\bf Noises } \end{sideways}} 
    &  Babble     & 0.53 & 0.53 & 0.54 & 0.55 & 0.56 & & 0.64 & 0.65 & 0.66 & 0.67 & 0.67 \\ \hhline{~------------} 
    &  Cafeteria  & 0.54 & 0.55 & 0.56 & 0.56 & 0.57 & & 0.65 & 0.66 & 0.67 & 0.67 & 0.68 \\ \hhline{~------------}
	&  Chainsaw   & 0.48 & 0.49 & 0.50 & 0.51 & 0.52 & & 0.57 & 0.58 & 0.59 & 0.61 & 0.62 \\ \hhline{~------------}
	&  SSN        & 0.52 & 0.52 & 0.53 & 0.54 & 0.55 & & 0.62 & 0.63 & 0.64 & 0.65 & 0.66 \\ \hline	
\end{tabular}
\label{Tab::ESII_UNP} 
\vspace{-.45cm}
\end{table}
\begin{table}[t!]
  \caption{ASII$_{ST}$ intelligibility measure [\%] for UNP speech signals}
  \newcommand{\mc}[3]{\multicolumn{#1}{#2}{#3}}
  \centering
  \renewcommand{\tabcolsep}{1.0mm}
  \vspace{-.2cm}
\begin{tabular}{|c|l c c c c c c c c c c c|} 	
   \hline
   \multicolumn{2}{|l}{ \bf  }    & \multicolumn{5}{c}{ Stairway ($T_{60} = 1.1$ s) } & & \multicolumn{5}{c|}{ Aula Carolina ($T_{60} = 4.9$ s) } \\ \hhline{~~-----~-----}
   \multicolumn{2}{|l}{ SNR (dB) }    & {\centering -3} & {\centering -2} & {\centering -1} & {\centering 0} & {\centering 1} & & {\centering -3} & {\centering -2} & {\centering -1} & {\centering 0} & {\centering 1} \\ \hline \hline
   \centering  \multirow{4}{*}{\begin{sideways} {\bf Noises } \end{sideways}} 
    &  Babble     & 0.58 & 0.59 & 0.60 & 0.61 & 0.61 & & 0.68 & 0.68 & 0.69 & 0.70 & 0.71  \\ \hhline{~------------} 
    &  Cafeteria  & 0.60 & 0.61 & 0.61 & 0.62 & 0.62 & & 0.69 & 0.70 & 0.70 & 0.71 & 0.71  \\ \hhline{~------------}
	&  Chainsaw   & 0.55 & 0.56 & 0.57 & 0.58 & 0.58 & & 0.62 & 0.63 & 0.64 & 0.65 & 0.66  \\ \hhline{~------------}
	&  SSN        & 0.58 & 0.58 & 0.59 & 0.60 & 0.61 & & 0.66 & 0.67 & 0.68 & 0.69 & 0.70  \\ \hline 
\end{tabular}
\label{Tab::ASII_UNP}
\end{table}
\begin{figure}[t!]
  \begin{center}
  \centering
  \vspace{-.3cm}
  \includegraphics[width=8.3cm]{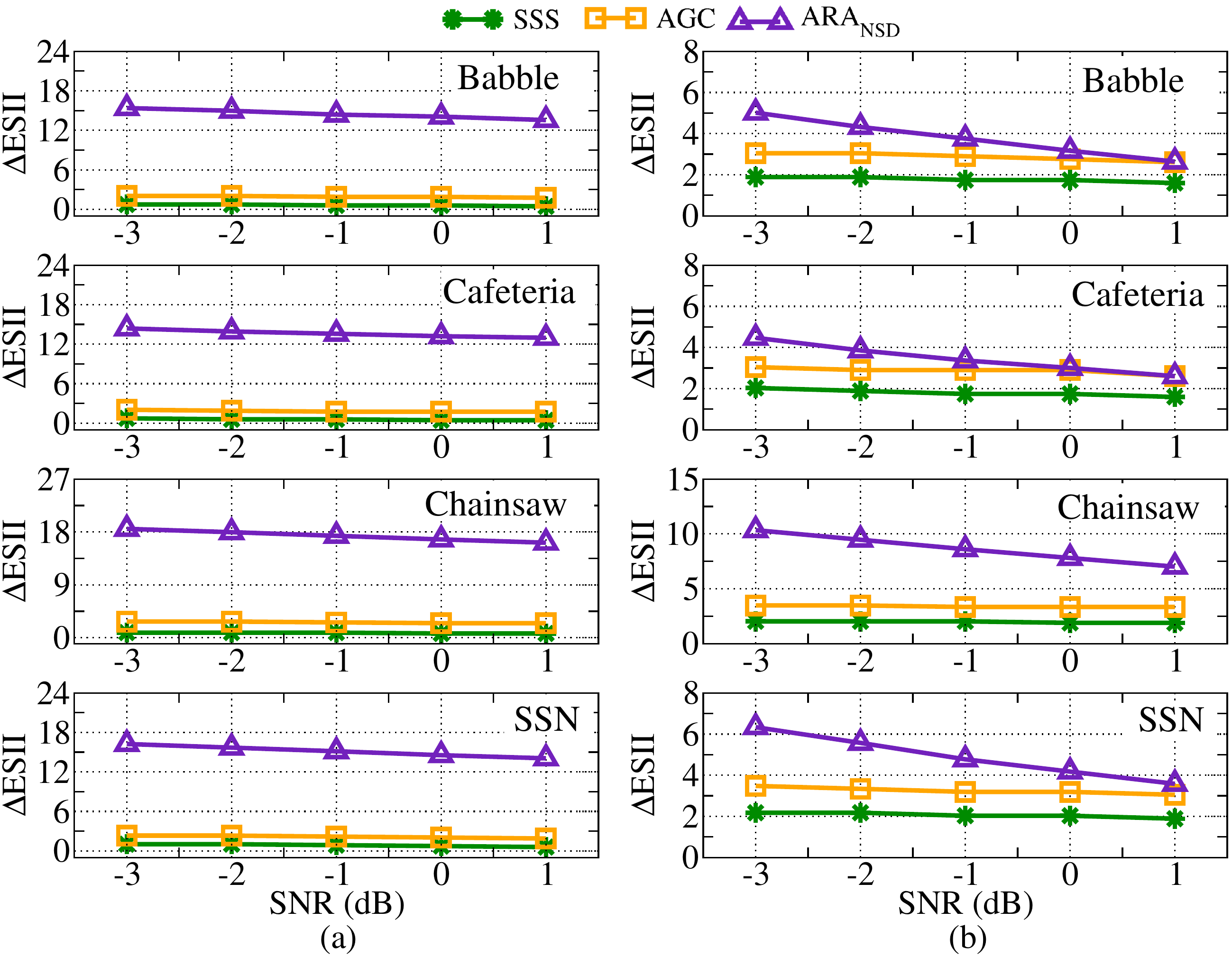}
  \vspace{-.3cm}
  \caption{$\Delta$ESII intelligibility improvement [{\tiny x$10^{-2}$}] for (a) the Stairway ($T_{60} = 1.1$ s) and (b) the Aula Carolina ($T_{60} = 4.9$ s) rooms.}
  \vspace{-0.9cm}
  \label{Fig::Delta_ESII}    		
  \end{center}
\end{figure}
The ESII \cite{Rheber_05} and ASII\textsubscript{ST} \cite{Hen_15} measures are adopted to evaluate the intelligibility improvement under non-stationary noisy-reverberant conditions. 
The direct path speech signal $s_{dir}(n)$ is chosen as the reference signal. 
Since late reverberation and additive background noise reduce intelligibility and are uncorrelated to $s_{dir}(n)$, the jointly distortion is obtained by the subtraction $s(n) - s_{dir}(n)$.  
These objective measures are normalized by the intelligibility achieved for the clean unprocessed signal corrupted by SSN noise at $20$ dB, considered here as a good intelligibility reference.

The ESII intelligibility improvement ($\Delta$ESII) is presented on Figure \ref{Fig::Delta_ESII} for the Stairway and Aula Carolina rooms.
In the first case, the ARA\textsubscript{NSD} outperforms the competing methods accomplishing more than seven times the AGC value in most of the cases.
For the Cafeteria scenario at $-1$ dB the proposed technique achieves an improvement of $13.6$, which corresponds to an assessment eight times the value of $1.7$ for the AGC method. 
As the Stairway room presents a DRR of $-9.1$ dB, the reverberation energy in this room is considerably higher than the energy related to the direct signal.
This means that the masking components are highlighted in this scenario.
As the proposed ARA\textsubscript{NSD} is an absorption approach designed to detect such effects, it is able to effectively reduce the temporal coloration.  
The SSS technique presents the smallest overall intelligibility improvement. 
Considering the Aula Carolina room, the proposed method also achieves the highest improvement for most of the cases.
This is observed for all noisy-reverberant conditions contemplating SNRs below or equal to $0$ dB.
The best $\Delta$ESII results are obtained by ARA\textsubscript{NSD} considering the most challenge condition of Chainsaw acoustic noise.	
The ARA\textsubscript{NSD} technique presents similar improvement as AGC for both Babble and Cafeteria noises at $1$ dB.
\begin{figure}[t!]
  \begin{center}
  \centering
  \includegraphics[width=8.3cm]{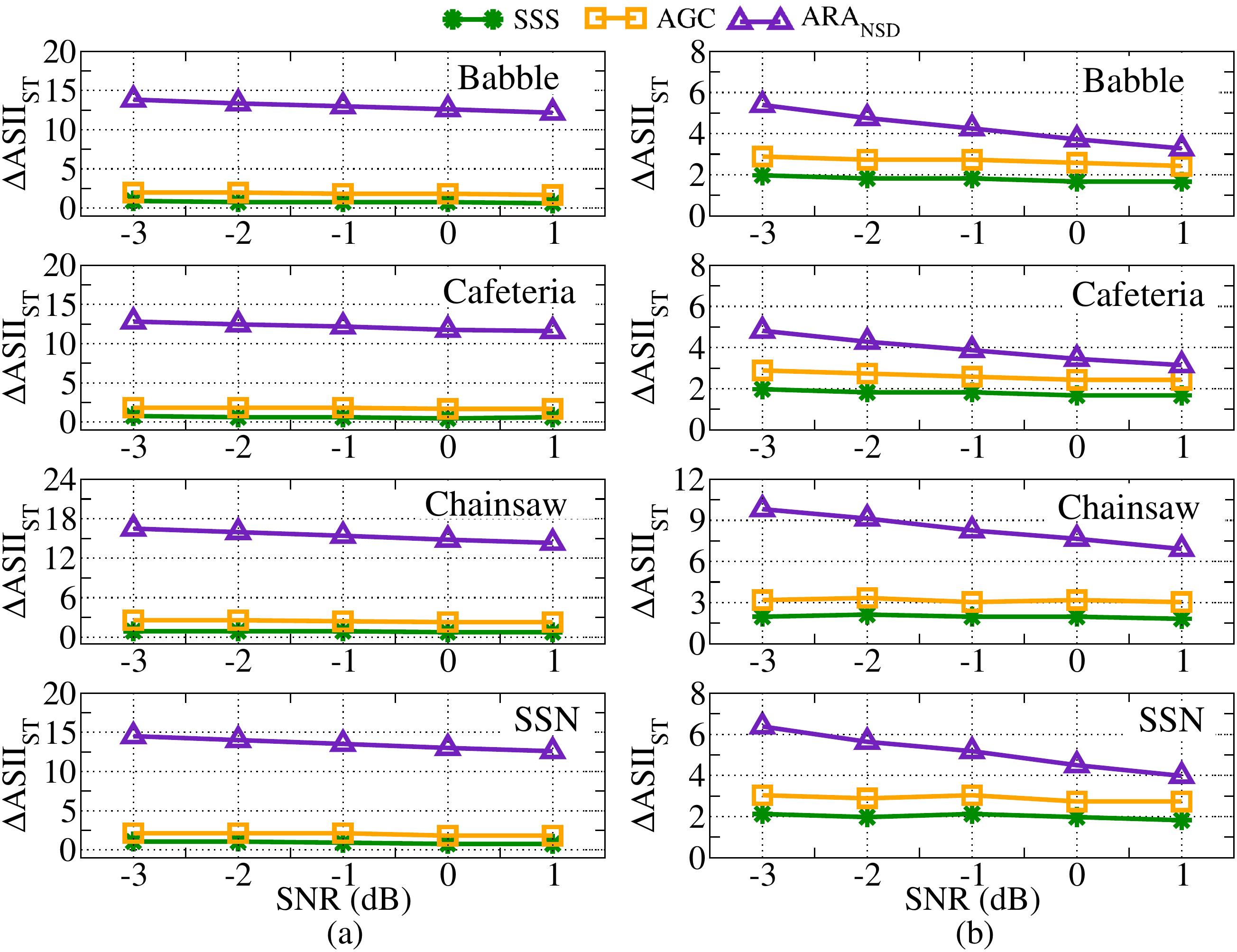}
  \vspace{-.3cm}
  \caption{$\Delta$ASII$_{\text{ST}}$ intelligibility improvement [{\tiny x$10^{-2}$}] for (a) the Stairway ($T_{60} = 1.1$ s) and (b) the Aula Carolina ($T_{60} = 4.9$ s) rooms.}
  \vspace{-.9cm}
  \label{Fig::Delta_ASII}    		
  \end{center}
\end{figure}

{\color{black} Figure \ref{Fig::Delta_ASII} depicts the $\Delta$ASII\textsubscript{ST} values for both reverberation rooms. 
For the Stairway room, the proposed method effectively attenuates masking components and attains the highest intelligibility improvement results for all conditions with $\Delta$ASII\textsubscript{ST} values above $10$.
The ARA\textsubscript{NSD} accomplished the highest overall $\Delta$ASII\textsubscript{ST} of $16.5$ for the highly non-stationary Chainsaw noise at $-3$ dB.     
Baseline technique  SSS is outperformed by the ARA\textsubscript{NSD} and AGC algorithms in all scenarios. 
The ARA\textsubscript{NSD} also presents the best $\Delta$ASII\textsubscript{ST} intelligibility results for the Aula Carolina room. 
Once again, the highly non-stationary Chainsaw noise leads to the most challenge condition.
In this case, the ARA\textsubscript{NSD} is still able to achieve an average improvement of $8.3$, compared to $3.6$ and $1.2$ for the AGC and SSS techniques, respectively. }



{\color{black} The SRMR quality metric \cite{Falk_10_2} estimates the human perceived reverberation effect on speech signals. 
Its updated version, the SRMR\textsubscript{norm} \cite{Falk_14}, is also selected for objective evaluation.}
The goal is to distinguish among the three approaches the ones that can better mitigate temporal coloration on speech signals.
The direct signal is used as a reference for normalization, such that the  SRMR\textsubscript{norm} presents values ranging between $[0,1]$, where $1$ determines a reverberation free signal. 

\begin{figure}[t!]
  \begin{center}
  \centering
  \includegraphics[width=8.5cm]{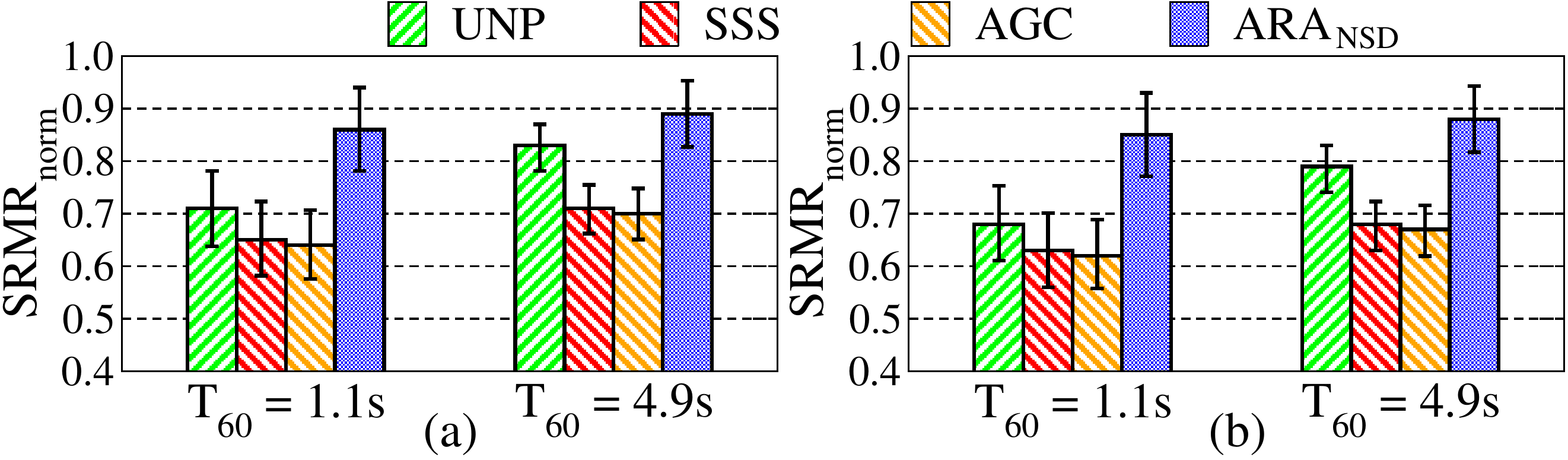}
  \vspace{-.4cm}
  \caption{SRMRnorm for (a) Noise-free reverberation and (b) reverberation with SSN at $0$ dB.}
  \label{Fig::SRMRnorm}    		
  \end{center}
\end{figure}

Figure \ref{Fig::SRMRnorm} illustrates the average SRMR\textsubscript{norm} values for the Stairway and Aula Carolina rooms under a noise-free reverberation condition (a) and a noisy-reverberation scenario with SSN background noise at $0$ dB (b).
Note that the ARA\textsubscript{NSD} attains the best SRMR\textsubscript{norm} values for all situations with a mean of $0.85$ and $0.89$ for the Stairway and Aula Carolina rooms, respectively.
This implies that the proposed method achieved an average quality increment of $0.16$ and $0.08$ for these rooms when compared with the UNP case. 
The SSS and AGC techniques present similar behavior, attaining the worst average SRMR\textsubscript{norm} values in these scenarios.
These results reinforce the capacity of the proposed method to absorb masking components {\color{black} providing intelligibility and quality  improvement}.

\begin{figure}[t!]
  \begin{center}
  \centering
  \vspace{-.5cm}
  \includegraphics[width=7cm]{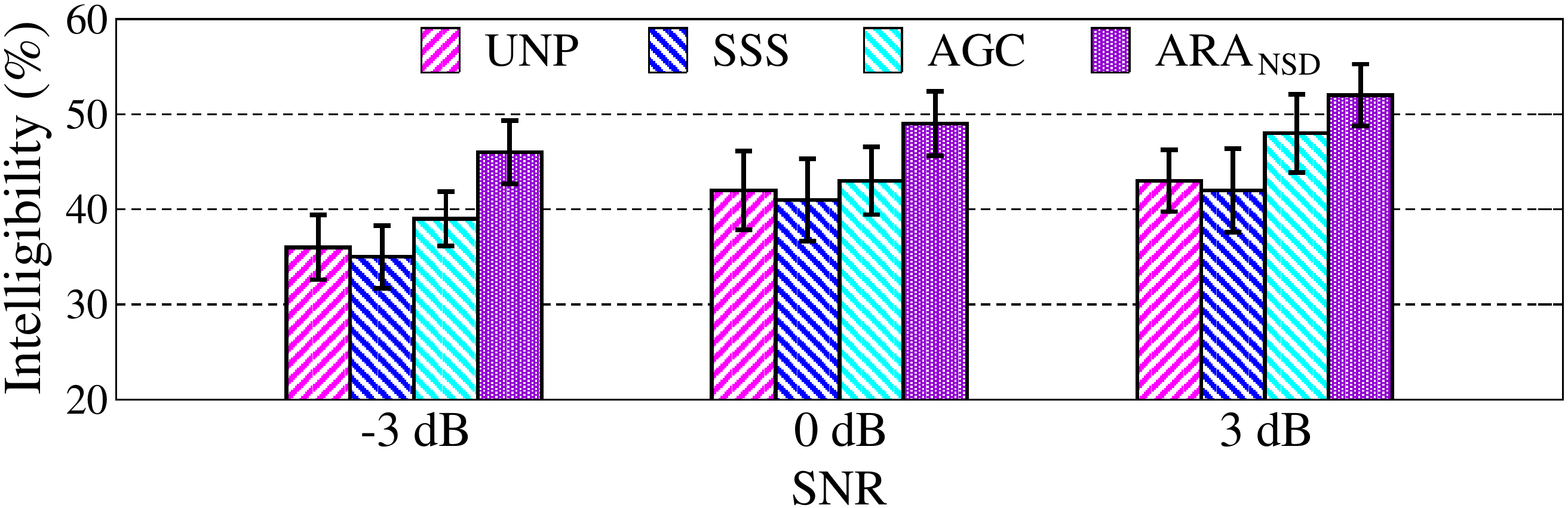}
  \vspace{-.4cm}
  \caption{Perceptual intelligibility evaluation for ISM room ($T_{60} = 1.0$ s) and SSN additive acoustic noise.}
  \vspace{-.75cm}
  \label{Fig::Percep}    		
  \end{center}
\end{figure}

\vspace{-.4cm}  
\subsection{Subjective Intelligibility Evaluation} 
\vspace{-.15cm} 
\color{black}
A listening test \cite{Ghimire_12} with ten native male Brazilian volunteers was conducted considering a closed scenario of phonetic balanced words\footnote{ The complete test database is available at lasp.ime.eb.br.}.
Their ages ranged from $22$ to $41$ years with an average of $32$.
A simulated room with $7.0$ x $5.2$ x $3.0$ m$^3$ and $T_{60}=1.0$ s was generated by the image source method (ISM) \cite{Allen_79}.
The SSN acoustic noise was adopted with SNRs of $-3$ dB, $0$ dB and $3$ dB.
Ten words were applied for each of $12$ test conditions, i.e., three SNR levels for three methods plus the unprocessed case.  
Participants were introduced to the task in a training session with 8 words.
The material was diotically presented using a pair of Roland RH-200S headphones. 
Listeners heard each word once in an arbitrary presentation order. 


The average intelligibility scores and standard deviations values for each method are presented in Figure \ref{Fig::Percep}.
The ARA\textsubscript{NSD} improves the intelligibility under all conditions over competing techniques. 
The proposed method improves $9$, $7$ and $9$ against $3$, $1$ and $5$ for the AGC under SNR values of $-3$ dB, $0$ dB and $3$ dB, respectively. 
In accordance with findings of \cite{Petkov_16}\cite{Hen_15}, SSS attains scores less than or equal to the UNP case. 


\color{black}
\vspace{-.35cm}  
\section{Conclusion}
\vspace{-.15cm} 
This letter proposed a new time domain absorption approach designed to reduce masking components of speech signals under noisy-reverberant conditions.
In this method, the non-stationarity of segments of the corrupted signal is used to detect masking distortions based on a defined threshold.
The non-stationarity degree was objectively measured with the INS and was also adopted to determine the absorption procedure.
Two reverberant rooms and four acoustic noises were used to compose the noisy-reverberant scenarios.
{\color{black} Two intelligibility measures were used for objective evaluation.
The results showed that the proposed scheme leads to a higher intelligibility improvement when compared to competing methods.
A perceptual listening test corroborated these results.  
An objective quality measure demonstrated that the proposed technique attains quality improvement.}



\bibliographystyle{ieeetr}
\newpage
\balance
\bibliography{SPL_2019_AfterR}

\end{document}